# Robust Feature Disentanglement in Imaging Data via Joint Invariant Variational Autoencoders: from Cards to Atoms

Maxim A. Ziatdinov[1,2,a] and Sergei V. Kalinin[1]

[1] Center for Nanophase Materials Sciences, Oak Ridge National Laboratory, Oak Ridge, TN 37831, USA

[2] Computational Sciences and Engineering Division, Oak Ridge National Laboratory, Oak Ridge, TN 37831, USA

Recent advances in imaging from celestial objects in astronomy visualized via optical and radio telescopes to atoms and molecules resolved via electron and probe microscopes are generating immense volumes of imaging data, containing information about the structure of the universe from atomic to astronomic levels. The classical deep convolutional neural network architectures traditionally perform poorly on the data sets having a significant orientational disorder, that is, having multiple copies of the same or similar object in arbitrary orientation in the image plane. Similarly, while clustering methods are well suited for classification into discrete classes and manifold learning and variational autoencoders methods can disentangle representations of the data, the combined problem is ill-suited to a classical non-supervised learning paradigm. Here we introduce a joint rotationally (and translationally) invariant variational autoencoder (*j-tr*VAE) that is ideally suited to the solution of such a problem. The performance of this method is validated on several synthetic data sets and extended to high-resolution imaging data of electron and scanning probe microscopy. We show that latent space behaviors directly comport to the known physics of ferroelectric materials and quantum systems. We further note that the engineering of the latent space structure via imposed topological structure or directed graph relationship allows for applications in topological discovery and causal physical learning.

[a] ziatdinovma@ornl.gov

Physical and chemical imaging tools are now the mainstay of multiple domain areas ranging from astronomy and medicine to materials science and condensed matter physics. On the macroscopic scales, optical and radio telescopes provide detailed maps of the universe. On a human scale, medical imaging and tomography are now an inseparable part of medical diagnostic and intervention.[1,2] Finally, techniques such as electron[3,4] and scanning probe[5] microscopies, optical and mass-spectrometry based chemical imaging,[6,7] and focused X-Ray[8] underpin condensed matter physics, chemistry, and materials sciences. Common for all these extremely disparate disciplines is the generation of the vast volumes of the multidimensional and multimodal data, with examples ranging from classical 3D current imaging tunneling spectroscopy (CITS) in scanning tunneling microscopy (STM),[9] force-volume imaging in atomic force microscopy (AFM)[10,11] and multidimensional spectroscopic data sets in Piezoresponse Force Microscopy,[12] electron energy loss spectroscopy (EELS) in scanning transmission electron microscopy (STEM),[13] 4D data sets in ptychographic imaging in STEM[14-16] and focused X-Rays, and the immense variety of hyperspectral imaging modes in optical and mass-spectrometric imaging.

Correspondingly, analysis of the imaging data becomes the key challenge across these disciplines. In case of astronomy, this involves identification and classification of stellar objects such as planets, stars, and galaxies. In high-resolution microscopies, it is identification of atoms, defects, molecules, and elementary building blocks of matter. Finally, in other imaging disciplines this can be faces, handwritten digits, or myriad of possible objects of interest. Throughout the decades, such analyses were based on combination of hand-crafted features and often very complex workflows requiring manual tuning to account for small changes in data-generation process, i.e., out of distribution drift.

The deep convolutional neural networks (DCNNs) allow an alternative paradigm for the analysis of the imaging and hyperspectral data, with multiple target applications ranging from supervised (classification and regression) to unsupervised learning (clustering and dimensionality reduction).[17] By the nature of their construction, the DCNNs allow for the equivariance in feature detection, i.e. they allow the detection of features independent of their relative position within the image plane. This addresses one of the central challenges in feature discovery. However, the second challenge – the presence of the possible rotational variants of the same object and the general lack of rotational information priors – remains unaddressed. For the classical DCNNs, a discovery of the rotational variant of the same object is a well-recognized challenge .[18] Likewise, for linear and manifold learning methods, the unmixing in a spatial domain will yield multiple classes corresponding to the sampling of existing rotational variants.

Similarly, very often the imaging data is formed by objects of multiple classes with a certain degree of variability within the class. The typical example is the paradigmatic hand-written digit MNIST database.[19] However, these examples abound across all disciplines, from the spiral and spherical galaxies in astronomy to molecules and defects in atomically resolved images. In supervised and semi-supervised learning, the a priori available labels are used to establish relevant classes and propagate them from a small number of known examples to the full data set. Similarly, while clustering methods are well suited for classification into discrete classes and manifold learning and variational autoencoders methods can disentangle representations of the data, the combined problem is ill-suited to the classical non-supervised learning paradigm.

Here we introduce the general class of *j*oint *t*ranslationally and *r*otationally invariant variational autoencoders (*j-tr*VAE) that are ideally suited to the solution of such problem and demonstrate their applicability to several synthetic and physics-based data sets. Previously, we have explored the special cases of this approach, namely rotational VAEs .[20] Here, we generalize this approach and implement it in the form of comprehensive open-source Python package.

The regular VAE consists of a generative model $p_\theta$ that reconstructs data **x** given a latent "code" **z** and an inference model $q_\varphi$ whose role is to perform an approximate (variational) inference of the true posterior distribution $p_\theta(\mathbf{z}|\mathbf{x})$. The generative and inference models (frequently referred to as "decoder" and "encoder) are parametrized by deep neural networks whose parameters are learned (jointly) by maximizing evidence lower bound (ELBO) with stochastic gradient descent.

Here, our goal is to categorize data containing orientational and/or positional disorder into different discrete classes and find relevant continuous factors of variation for each class at the same time. However, the standard VAE implementation does not allow for categorical latent variables as they cannot be backpropagated through samples. One possible workaround for performing a joint (discrete + continuous) variational inference is to replace a non-differentiable sample from a categorical distribution with a differentiable sample from a Gumbel-Softmax distribution whose "temperature" parameter controls how closely it approximates the categorical distribution.[21, 22] This method, however, introduces bias and modeling assumptions. Here, instead, we marginalize out discrete latent variables *y* via a fully parallel enumeration replacing the Monte Carlo expectation $\mathbb{E}_{y \sim q_\varphi(\cdot|\mathbf{x})} \nabla \text{ELBO}$ with an explicit sum $\sum_y q_\varphi(y|\mathbf{x}) \nabla \text{ELBO}$. This approach allows achieving accurate results while using significantly fewer parameters in the inference and generator networks compared to the method based on the Gumbel-Softmax.

The next key element is the enforcement of geometrical consistency between rotated and/or shifted images. This is achieved by creating "special" latent dimensions designated to absorb the information about arbitrary rotations and translations, whose latent vectors **γ** and **s** are sampled from von Mises and Normal distributions, respectively. The sampled values are used to perform an affine transformation of the coordinate grid defined as $R(\boldsymbol{\gamma})\mathbf{x} + \mathbf{s}$, and the transformed grid is passed to a decoder (which is now a function of pixel coordinates[23]) together with the VAE continuous and discrete latent variables (see the details of implementation for convolutional and MLP decoders in the Supplementary Materials).

Overall, we define the generative process as following:

$$p(\mathbf{z}) = \mathcal{N}(\mathbf{z}|0, \mathbf{I}); \quad p(y) = Cat(y|\pi); \tag{1a}$$

$$p(\boldsymbol{\gamma}) = vM(\boldsymbol{\gamma}|k); \quad p(\mathbf{s}) = \mathcal{N}(\mathbf{s}|0, \sigma_\mathbf{s}); \tag{1b}$$

$$p_\theta(\mathbf{x}|\mathbf{z}, \mathbf{s}, \boldsymbol{\gamma}, y) = Bernoulli(\mathbf{x}|\mu(\mathbf{z}, \mathbf{s}, \boldsymbol{\gamma}, y)), \tag{1c}$$

where $p(\mathbf{z})$ is a standard normal prior for the continuous latent code, $Cat(y|\pi)$ is a categorical distribution associated with a (unknown) class label, $vM(\boldsymbol{\gamma}|k)$ is von Mises prior with concentration $k$ for the latent angle, $p(\mathbf{s})$ is a normal prior for the offsets, and $Bernoulli(\mathbf{x}|\mu(\mathbf{z}, \mathbf{s}, \boldsymbol{\gamma}, y))$ is a parametrized Bernoulli likelihood function with $\mu$ being a decoder neural network (with an affine transformation layer as an "input layer").

Our inference model is a multi-head neural network that outputs parameters of the distributions for continuous (**z**, **s**, **γ**) and discrete (*y*) latent variables. The objective "loss" function (the negative ELBO) for the *j-tr*VAE model can be written in terms of the reconstruction error (*RE*) and Kullback-Leibler divergence terms ($D_{KL}$) as following:

$$\mathcal{L}(\mathbf{x}) = RE + \beta_1(t)D_{KL}\big(q(\tau|\mathbf{x})\|p(\tau)\big) + \beta_2 D_{KL}\big(q(\mathbf{z}|\mathbf{x})\|p(\mathbf{z})\big) + \beta_3(t)D_{KL}\big(q(y|\mathbf{x})\|p(y)\big), \tag{2}$$

where $D_{KL}(q(\tau|\mathbf{x})\|p(\tau)) = D_{KL}(q(\mathbf{s}|\mathbf{x})\|p(\mathbf{s})) + D_{KL}(q(\pmb{\gamma}|\mathbf{x})\|p(\pmb{\gamma}))$ and $\beta_i(t)$ are time-dependent scale factors (during the model training the role of time is played by the "epoch" number). Here we set $\beta_1$ to the constant value of 1 throughout the training. We found empirically that optimal results are typically achieved by starting with high values of ($\beta_2$, $\beta_3$), then linearly decreasing them to much smaller values during the first ~20% of the training and keeping them constant for the remaining ~80%. Finally, to encourage a better categorization of data, at the beginning of training, the penalty on the $D_{KL}$ term for continuous latent variables ($\beta_2$) is chosen to be several times larger than for the discrete one ($\beta_3$).

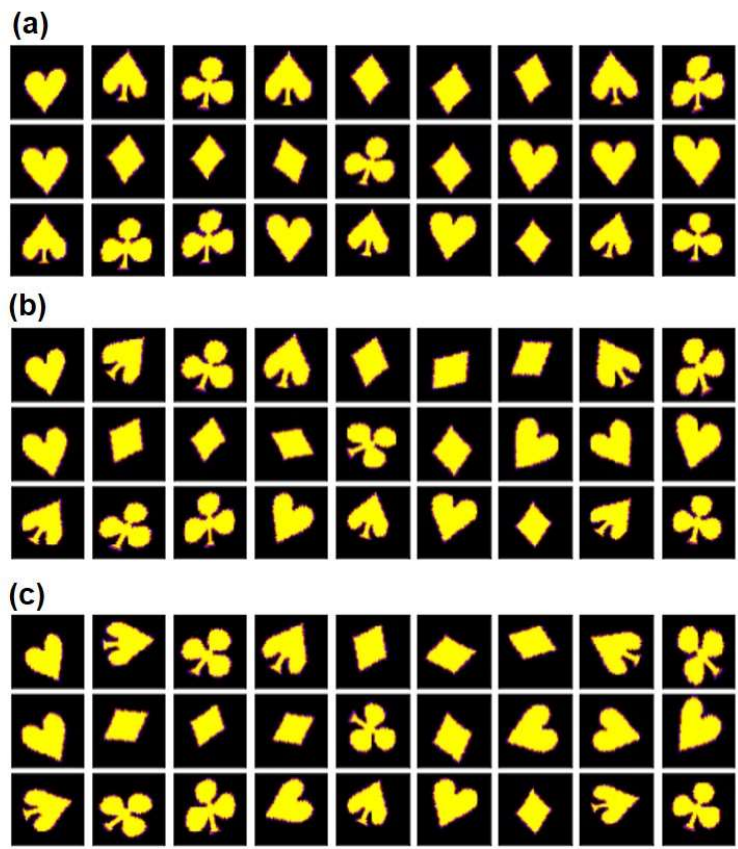

**Figure 1. Typical objects from the cards dataset for different degrees of orientational disorder.** (a) $\alpha \in [-12°, 12°]$, (b) $\alpha \in [-60°, 60°]$, (c) $\alpha \in [-120°, 120°]$. In all three cases a range of arbitrary translations is [0, 0.1] in the fractions of an image, a shear range is [-10, 10], and a scale range is [1, 1.2].

We start by demonstrating the application of *j-tr*VAE to a synthetic dataset with a known ground truth. Specifically, we create our dataset from playing card suits, with monochrome clubs, spades, diamonds, and hearts. We then apply a set of affine transforms including rotation, shear, translation, and scale to augment the dataset. We argue that the cards dataset offers an ideal toy model and can be an alternative to the dSprites dataset.[24] Here, diamonds have a shape that is very distinct from the other three hands. Their rotation by 90 degrees is equivalent to uniaxial compression and resizing, bringing interesting degeneracy into outcomes of possible affine transforms. Similarly, hearts and spades differ by fairly small detail, whereas spades and clubs (without tail) have three-fold and mirror symmetry respectively. In the following, we focus on three sets with similar translational and shape (shear, scale) disorders but a drastically different rotational disorder as depicted in Figure 1.

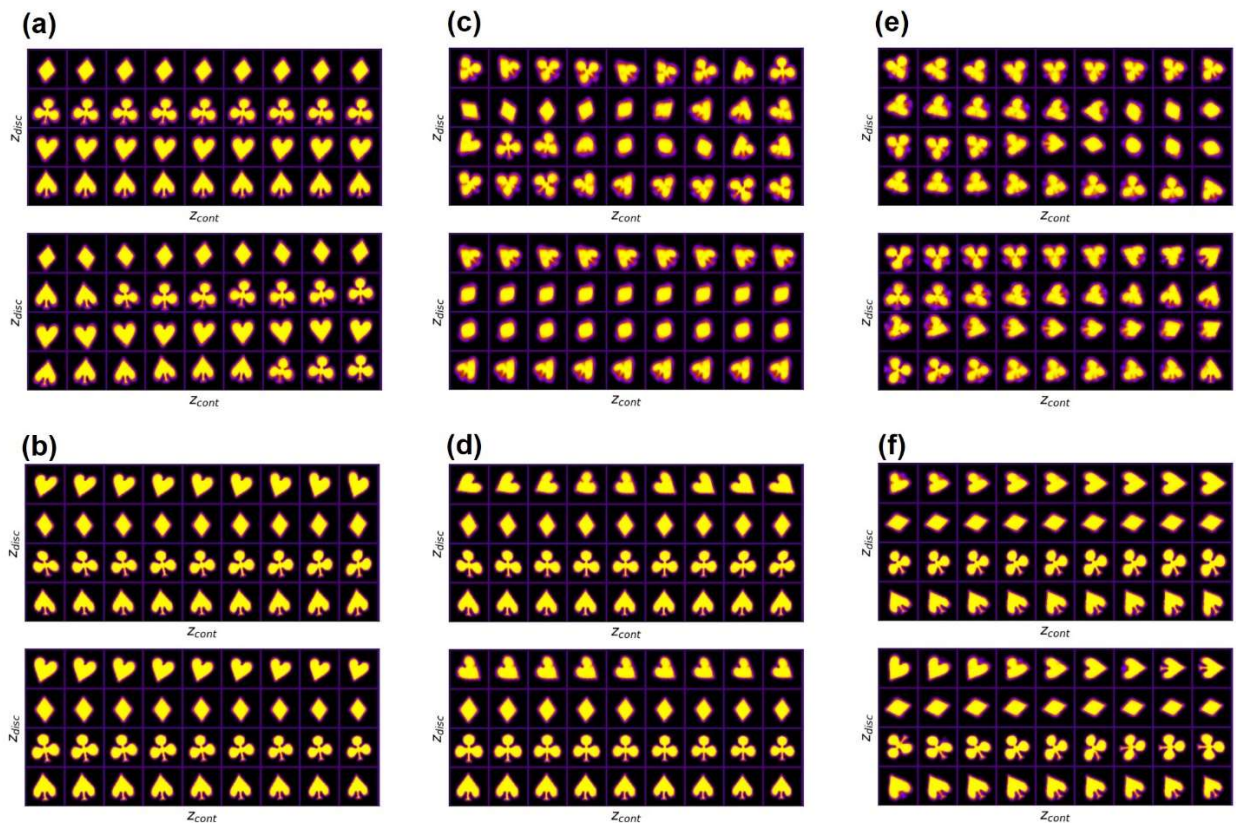

**Figure 2. Traversals of the learned latent manifolds of *j*-VAE and *j-tr*VAE for the cards dataset.** (a, c, e) *j*-VAE results for three different ranges of orientational disorders (small to large, see Fig. 1). (b, d, f) *j-tr*VAE results for the same data. In all the subfigures, the top and bottom subplots correspond to first and second continuous latent dimension, respectively, whereas the vertical ($z_{disc}$) and horizontal ($z_{cont}$) directions in each individual plot correspond to the discrete and continuous latent variables, respectively. The [min, max] range for the first (second) continuous latent variable is defined by sweeping over linearly spaced coordinates transformed through the inverse CDF of a Gaussian, whereas the second (first) continuous latent variable is set to a fixed value of 0. See Supplemental Materials for the plots in 2D latent space of the continuous variables for each identified class.

The traversals of the learned latent manifolds of *j*-VAE and *j-tr*VAE for the cards dataset are shown in Figure 2. In all the cases, the number of discrete and continuous latent dimensions was set to four and two, respectively. Here the vertical direction corresponds to the learned discrete classes whereas the horizontal direction is associated with the continuous factors of the variation. One can see that while the performance of the *j*-VAE can be compared to that of the *j-tr*VAE for the small orientational disorder (Fig. 2a, b), it drastically deteriorates for moderate and strong disorders (Fig. 2c, e) where neither discrete classes nor continuous traits can be recovered from the data. On the other hand, *j-tr*VAE clearly can separate different discrete classes and find relevant continuous factors of variation in each class. Indeed, for small and moderate orientational disorder Fig. 2(b, d) the first continuous latent variable (top subplots) corresponds to the variation in the shear strength whereas the second one (bottom subplots) is clearly associated with the scale variation ("zooming-out" from left to right). For the large orientational disorder (Fig. 2f), the disentanglement of continuous factors of variation is less clean since the *j-tr*VAE model wasn't able to account for 120° degenerate minima. In Figure 3 we show a correlation matrix analysis of the inferred discrete classes *versus* the ground truth labels for the moderate and strong disorder. We can see that overall, a model performed remarkably well separating different classes without supervision, while also discovering continuous traits for each class.

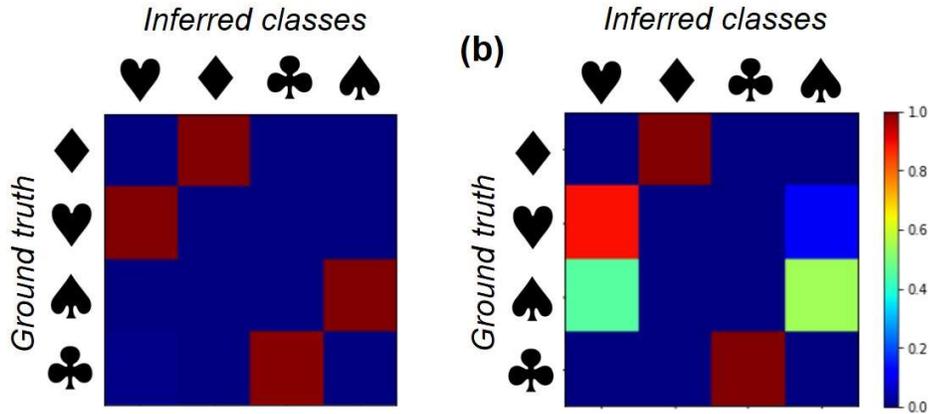

**Figure 3.** Correlation matrix analysis (inferred *vs.* ground truth). (a) moderate orientational disorder ($\alpha \in [-60°, 60°]$), (b) strong orientational disorder ($\alpha \in [-120°, 120°]$).

The remarkable property of the *j-tr*VAE is that it is possible to condition our generative process on both discrete class label and rotation angle (as well as a lateral shift) to generate a rotation of an object that was *not* in the initial training set. This is illustrated in Fig. 4 (a) where we used a model trained on the dataset with a small orientational disorder ($\alpha \in [-12°, 12°]$) to generate two orientations differed by 180° for a single class from different quadrants of the 2D latent space. Similarly, one can populate different parts of the latent space with different classes rotated by a specific angle as illustrated in Fig. 4 (b).

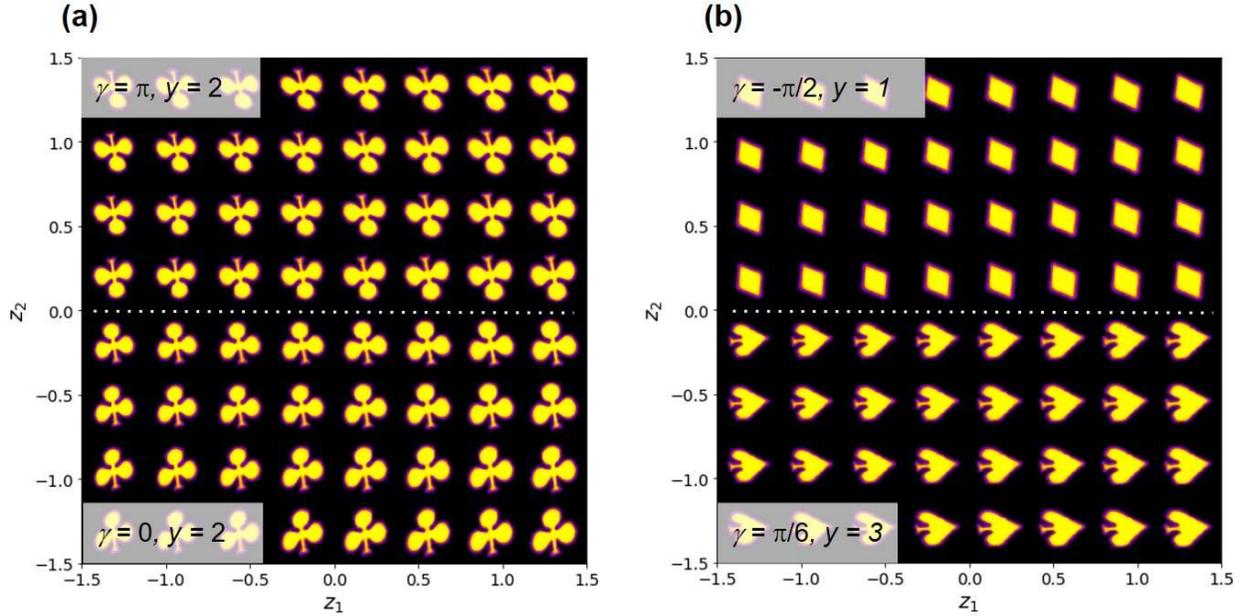

**Figure 4.** Class-conditioned and angle-conditioned data generation with a trained *j-tr*VAE model. The angle and class on which a generative model is conditioned are shown in the insets for different parts of the latent space.

We further illustrate applications of the *j-tr*VAE for several physical problems from classical and quantum systems. We note that in many cases the direct image-by-image analysis of the collection of images is intractable and hence the first step of analysis is the creation of the feature set. If no prior information is available, this can be based on the rectangular grids. In this case, the features represent image patches (2D), multilayer patches (3D), and spectral patches (3D and 4D voxels) of the initial dataset. Alternatively, the feature set can be defined using patches centered at the specific feature of interest, e.g., atoms in atomically-resolved images and spectroscopies, particles center of gravity, etc. While not yet explored in the context of the physical data analysis, more complex positional encoding schemes can be envisioned analogously with the attention-based natural language processing and image analysis models. Note that this choice introduces significant bias on the analysis and at the same time optimizes it for specific domain applications, somewhat equivalently on the constraints in the linear unmixing methods.

Here, we demonstrate that a *j-tr*VAE can find physically-meaningful discrete and continuous latent representations in the unsupervised manner from scanning transmission electron microscopy (STEM) data on a multiferroic material and scanning tunneling microscopy (STM) data from graphene. Both these systems have been extensively explored before and we refer the reader to the original publications on BFO[25-27] and graphene[28] for sample preparation and imaging details.

Figure 5a shows an atomically-resolved STEM image of paradigmatic multiferroic material BiFeO$_3$ (BFO) on SrTiO$_3$ (001) substrate. The image represents a 2D projection of a 3D material structure. The spatial degeneracy of the ferroelectric polarization commonly results in the formation of multiple nanoscale domains,[29] which can be identified "manually" from the relative positions of atomic centers in each unit cell (shown in the inset of Fig. 5a). The question that we ask is whether *j-tr*VAE can learn the spatial distribution of the polarization from the raw image data while separating the substrate and ferroic material into different classes, all in the unsupervised fashion.

To create a training set for the *j-tr*VAE, we extracted small image patches around the centers of the masses of individual unit cells. For each image patch, we store its position in the original image so that we can map the encoded values back to the real space. During the model training, we turned off the offset latent variable (i.e., no translational invariance) since in this case shifts may be relevant to the system's behavior and should be encoded in the "structural" latent variables.

The *j-tr*VAE results are shown in Figure 5b-d. Here we encoded each image patch using the trained inference model and projected the encoded values back onto the original image. By setting the discrete dimension to two, we were able to easily separate the substrate from the BFO material (Fig. 5b,c). For the class associated with BFO material, the learned latent manifold (Fig. 5b) reveals a variation in the relative position of the central atom with respect to atoms in the corners, which corresponds to the changes in the polarization value. Indeed, the distribution of the continuous latent variables (Fig. 5d,e) shows a good correspondence with the manually computed distributions for *x* and *y* components of the polarization (see Supplemental Materials). Hence, our model was able to learn a physically meaningful representation (ferroic variants) from atomically-resolved data in the unsupervised fashion.

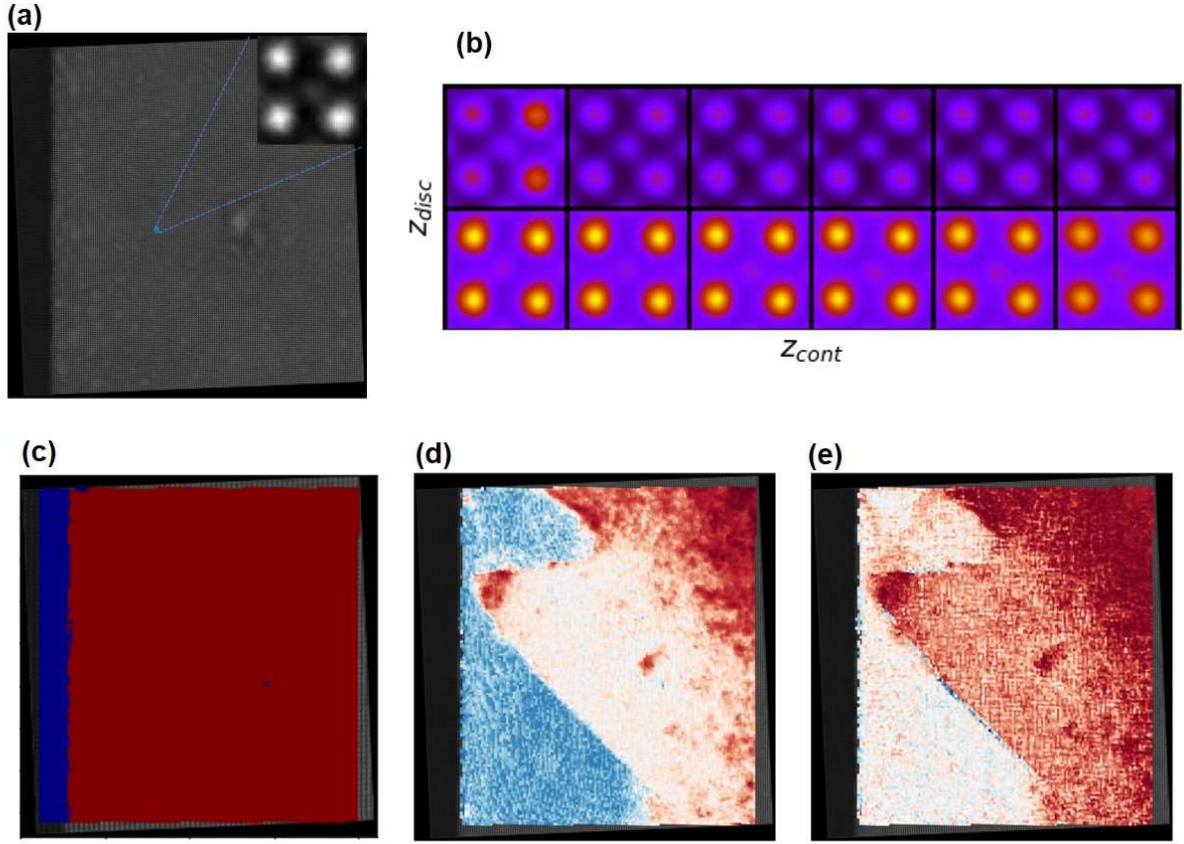

**Figure 5. Application of *j-tr*VAE to analysis of atomically-resolved structural data from a multiferroic material.** (a) STEM image of BFO film (on the right) on a substrate (on the left). Inset shows a magnified individual atomic unit cell (2D projection). The relative positions of atoms (4 brighter blobs in the corners and one darker blob around the center) in each unit cell determines a local polarization value. (b) The traversal of the learned latent manifold for the first continuous latent variable (for the second one, which is available in Supplementary Materials, the contrast variations are harder to discern). (c) Spatial distribution of discrete classes inferred by *j-tr*VAE. (d, e) Spatial distribution of the continuous latent variables for the discrete class associated with the BFO material.

As a second example, we apply the *j-tr*VAE model to quantum interference patterns in a graphene material observed in STM (Figure 6a). Here, point-like defects lead to the large momentum scattering of the Dirac electrons.[30-32] The (back-)scattered electron waves interfere with the incoming ones and produce a characteristic standing wave pattern whose period differs by $\sim\sqrt{3}$ from that of the underlying graphene lattice. However, the exact symmetry of the produced interference patterns strongly depends on the relative position and type of defects. Our goal is to separate a clean graphene area from the regions dominated by quantum interference patterns. In addition, we would also like to identify the positions of the defects themselves.

Similar to the analysis of STEM data, we create a training set by using patches of images. In the case of STM, however, we do not usually have (reliable) information about the atomic positions. Hence, we extract image patches using a uniform 2D grid of points. Next, instead of working with real-space data, we use the Fourier transformed (FT) data, which is a convenient way to analyze electron scattering patterns in STM.[33, 34] Finally, the FT patches are used as the inputs for the *j-tr*VAE training. The model is trained without the offset latent variables since the FT data is centered by the definition.

Figure 6(b,c) shows the traversals of the learned latent manifold for the two continuous latent variables where one of the discrete dimensions (top rows in 6b and 6c) can be immediately recognized as graphene lattice without quantum interference patterns. The projection of the encoded values to the real space is shown in Fig. 6d for the discrete latent variable. It becomes clear that the remaining two discrete latent dimensions are associated with defects and interference patterns occurring in-between the defects. We note, however, that both the latent manifold and spatial distribution of the encoded data suggest that there is a "leakage" from the second to third component and *vice versa*. This is not surprising as the defects are the *cause* of the observed interference patterns. Figure 6(e, f) shows the spatial distribution of the encoded continuous latent variables for the discrete latent variable associated with defects. Perhaps not surprisingly, the continuous variables show maxima/minima at the defects themselves. It is noteworthy, however, that the first continuous latent variable takes a drastically different value on the "outlier" defect which has likely a different structure.

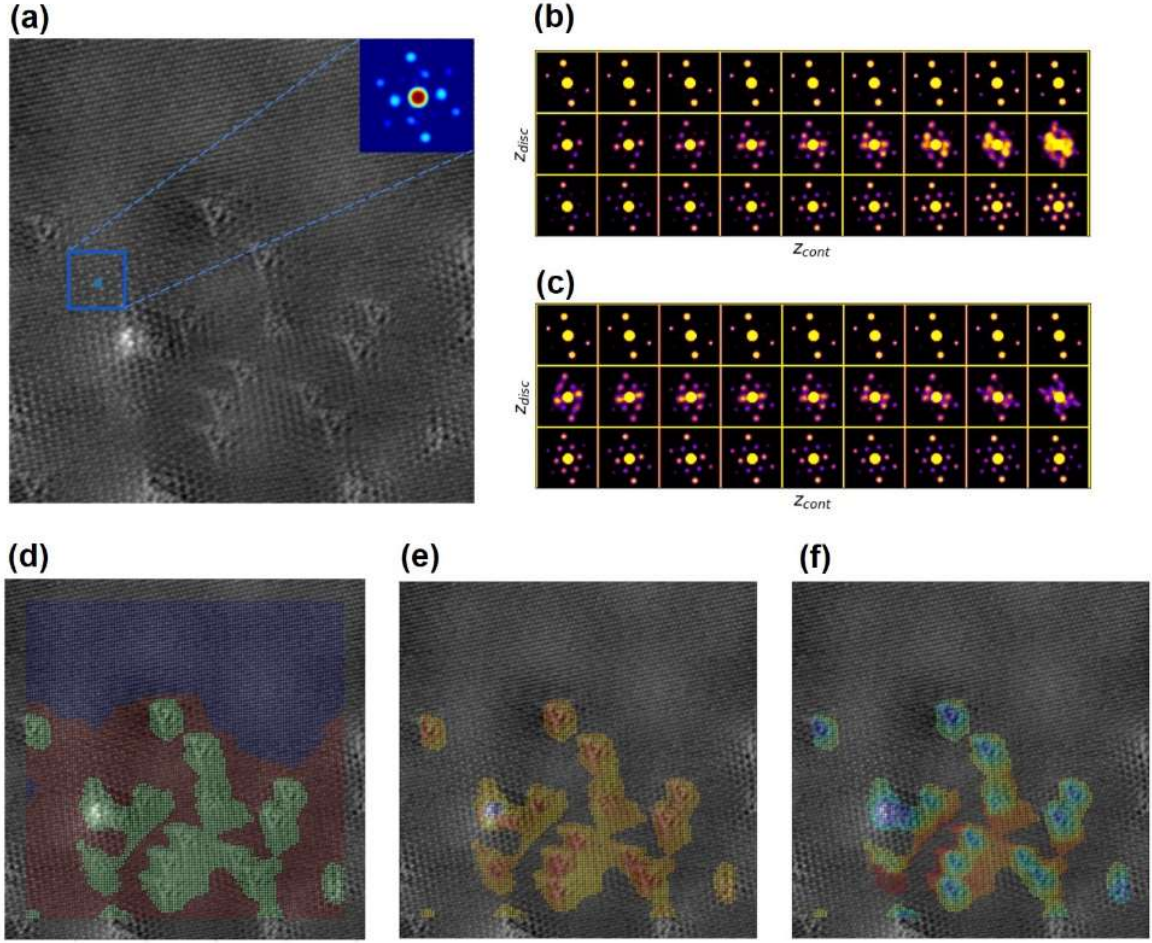

**Figure 6. Application of *j-tr*VAE to analysis of quantum interference patterns in graphene.** (a) Experimental STM image of graphene with point defects. Note that the upper part of the image shows a relatively clean graphene area where an atomic lattice can be seen. The inset shows a Fourier transformed image patch. (b, c) Traversals of the learned latent manifold for the first (b) and second (c) continuous latent dimensions. (d) Projection of the encoded discrete latent values onto the real space. (e, f) Projection of the encoded continuous latent values onto the real space for the first (e) and second (f) continuous latent dimensions for the discrete class associated with the defects.

We also demonstrate that it is possible to use a trained inference model of the *j-tr*VAE to locate defects in other STM data potentially providing an important ingredient for the autonomous experimentation.[35-38] Shown in Figure 7a is an STM scan obtained under same imaging conditions from a different sample region containing multiple new types of point defects. The real-space projection of the discrete and continuous latent variables encoded using the pre-trained inference model is shown in Fig.7b,c. Clearly, the model was able to identify regions with point defects, although the localization area is broader compared to the original data in Fig. 6. Nevertheless, this opens a path toward improving the generalizability by tuning the neural network architectures, diversifying the initial training set (currently, it was trained only on a single STM scan), etc.

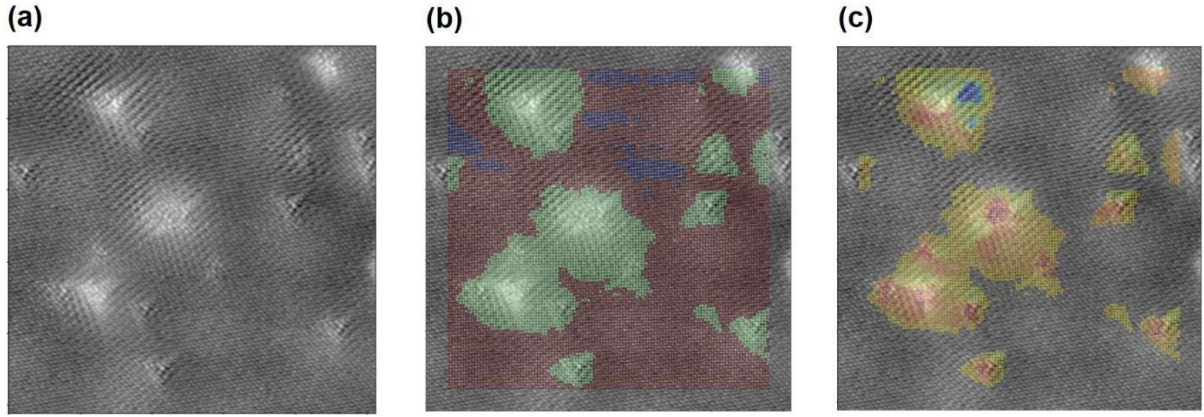

**Figure 7. Application of the pre-trained inference network to new data characterized by a distributional shift.** (a) Experimental STM image of graphene with point defects from a different sample region. Note that here we have several new types of larger defects compared to the training data from Fig. 6, i.e., there is a distribution shift. (b, c) Projection of the encoded discrete (b) and continuous (c) latent values onto the real space. Only the first continuous latent variable is shown, since the second one does not add new information.

To summarize, we developed and implemented joint rotationally and translationally invariant variational autoencoders, providing an approach to simultaneously categorize and disentangle the latent representations of imaging data in the presence of strong rotational disorder. We believe this to be a universal method for the discovery of the spatial features in multidimensional and multimodal data sets that can be applicable across multiple disciplines. Here, this approach is demonstrated for a synthetic data set and two examples from the field of scanning transmission electron microscopy and scanning tunneling microscopy respectively. However, the approach and the code are universal and can be used for any imaging data set.

We further note several important opportunities opened by this development. The first is related to the topological structure of the latent space. Here, we implemented rotationally and shift-invariant VAE, equivalent to the SO(2) Lie group. However, the generalization to other Lie groups as necessary for the data with known topological properties or topological structure discovery is straightforward[39, 40] and will be explored further. Similarly, the VAE architecture allows encoding the directed graph relationship between the variables[41], enabling applications in physical discovery and causal structure analysis.

**Methods:**

The *j-tr*VAE was realized in Pyro probabilistic programming language.[42] For all the case studies presented in the paper, both generative (decoder) and inference (encoder) models were parametrized by the multi-layer perceptrons (MLPs). The decoder was a 3-layer MLP with 256 hidden units ("neurons") in each layer activated by a hyperbolic tangent. The encoder was a 2-layer MLP with 256 neurons per layer activated by the hyperbolic tangent. The decoder and encoder weights were optimized jointly via Adam optimizer[43] with a learning rate between 0.0005 and 0.001. In all the cases we used 2 continuous latent variables. We found that in practice a better *j-tr*VAE convergence can be achieved by approximating the von-Mises distribution in Eq. (1b) with a Projected Normal distribution,[44] which is qualitatively similar but permits tractable variational inference via reparametrized gradients. The code will be made available at https://github.com/ziatdinovmax/jtrVAE

**Data Availability:**

The multiferroic data is available without restrictions from http://doi.org/10.5281/zenodo.4555979

**Acknowledgements:**

This work was performed and supported (M.Z.) at Oak Ridge National Laboratory's Center for Nanophase Materials Sciences (CNMS), a U.S. Department of Energy, Office of Science User Facility and partially supported (S.V.K.) by the U.S. Department of Energy (DOE), Office of Science, Basic Energy Sciences (BES), Materials Sciences and Engineering Division.

# References

1. S. J. Wang and R. M. Summers, Med. Image Anal. **16** (5), 933-951 (2012).
2. F. Cichos, K. Gustavsson, B. Mehlig and G. Volpe, Nat. Mach. Intell. **2** (2), 94-103 (2020).
3. S. J. Pennycook and P. D. Nellist,  (Springer, New York, 2011).
4. B. L. Mehdi, M. Gu, L. R. Parent, W. Xu, E. N. Nasybulin, X. L. Chen, R. R. Unocic, P. H. Xu, D. A. Welch, P. Abellan, J. G. Zhang, J. Liu, C. M. Wang, I. Arslan, J. Evans and N. D. Browning, Microsc. microanal. **20** (2), 484-492 (2014).
5. C. Gerber and H. P. Lang, Nat. Nanotechnol. **1** (1), 3-5 (2006).
6. D. J. Comstock, J. W. Elam, M. J. Pellin and M. C. Hersam, Analytical Chemistry **82** (4), 1270-1276 (2010).
7. O. S. Ovchinnikova, M. P. Nikiforov, J. A. Bradshaw, S. Jesse and G. J. Van Berkel, ACS Nano **5** (7), 5526-5531 (2011).
8. S. O. Hruszkewycz, C. M. Folkman, M. J. Highland, M. V. Holt, S. H. Baek, S. K. Streiffer, P. Baldo, C. B. Eom and P. H. Fuoss, Appl. Phys. Lett. **99** (23), 232903 (2011).
9. A. Asenjo, J. M. Gomezrodriguez and A. M. Baro, Ultramicroscopy **42**, 933-939 (1992).
10. A. Schirmeisen and B. Roling, Monatshefte Fur Chemie **140** (9), 1103-1112 (2009).
11. M. Gad, A. Itoh and A. Ikai, Cell Biology International **21** (11), 697-706 (1997).
12. S. Jesse, R. K. Vasudevan, L. Collins, E. Strelcov, M. B. Okatan, A. Belianinov, A. P. Baddorf, R. Proksch and S. V. Kalinin, Annual Review of Physical Chemistry, Vol 65 **65**, 519-536 (2014).
13. M. Bosman, M. Watanabe, D. T. L. Alexander and V. J. Keast, Ultramicroscopy **106** (11-12), 1024-1032 (2006).
14. S. Jesse, M. Chi, A. Belianinov, C. Beekman, S. V. Kalinin, A. Y. Borisevich and A. R. Lupini, Sci Rep **6** (2016).
15. C. Ophus, Microsc. microanal. **25** (3), 563-582 (2019).
16. Y. Jiang, Z. Chen, Y. M. Hang, P. Deb, H. Gao, S. E. Xie, P. Purohit, M. W. Tate, J. Park, S. M. Gruner, V. Elser and D. A. Muller, Nature **559** (7714), 343-+ (2018).
17. Y. LeCun, Y. Bengio and G. Hinton, Nature **521** (7553), 436-444 (2015).
18. T. S. Cohen and M. Welling, in *Proceedings of the 33rd International Conference on International Conference on Machine Learning - Volume 48* (JMLR.org, New York, NY, USA, 2016), pp. 2990–2999.
19. L. Deng, IEEE Signal Processing Magazine **29** (6), 141-142 (2012).
20. S. V. Kalinin, O. Dyck, A. Ghosh, B. G. Sumpter and M. Ziatdinov, arXiv preprint arXiv:2010.09196 (2020).
21. E. Jang, S. Gu and B. Poole, arXiv preprint arXiv:1611.01144 (2016).
22. E. Dupont, arXiv preprint arXiv:.00104 (2018).
23. T. Bepler, E. Zhong, K. Kelley, E. Brignole and B. Berger, Advances in Neural Information Processing Systems, 15409-15419 (2019).
24. https://deepmind.com/research/open-source/dsprites-disentanglement-testing-sprites-dataset.
25. C. T. Nelson, R. K. Vasudevan, X. H. Zhang, M. Ziatdinov, E. A. Eliseev, I. Takeuchi, A. N. Morozovska and S. V. Kalinin, Nat. Commun. **11** (1), 12 (2020).
26. M. Ziatdinov, C. T. Nelson, X. H. Zhang, R. K. Vasudevan, E. Eliseev, A. N. Morozovska, I. Takeuchi and S. V. Kalinin, npj Comput. Mater. **6** (1), 9 (2020).
27. M. Ziatdinov, C. Nelson, R. K. Vasudevan, D. Y. Chen and S. V. Kalinin, Appl. Phys. Lett. **115** (5), 5 (2019).
28. M. Ziatdinov, S. Fujii, M. Kiguchi, T. Enoki, S. Jesse and S. V. Kalinin, Nanotechnology **27** (49) (2016).
29. L. E. C. A.K. Tagantsev, and J. Fousek, *Domains in Ferroic Crystals and Thin Films*. (Springer, New York, 2010).

30. K. S. Novoselov, A. K. Geim, S. V. Morozov, D. Jiang, M. I. Katsnelson, I. V. Grigorieva, S. V. Dubonos and A. A. Firsov, Nature **438** (7065), 197-200 (2005).
31. A. H. Castro Neto, F. Guinea, N. M. R. Peres, K. S. Novoselov and A. K. Geim, Rev. Mod. Phys. **81** (1), 109-162 (2009).
32. B. Keimer and J. E. Moore, Nat. Phys. **13** (11), 1045-1055 (2017).
33. C. Gutiérrez, C.-J. Kim, L. Brown, T. Schiros, D. Nordlund, Edward B. Lochocki, K. M. Shen, J. Park and A. N. Pasupathy, Nature Physics **12** (10), 950-958 (2016).
34. P. Roushan, J. Seo, C. V. Parker, Y. S. Hor, D. Hsieh, D. Qian, A. Richardella, M. Z. Hasan, R. J. Cava and A. Yazdani, Nature **460** (7259), 1106-U1164 (2009).
35. A. Krull, P. Hirsch, C. Rother, A. Schiffrin and C. Krull, Commun. Phys. **3** (1), 8 (2020).
36. R. K. Vasudevan, K. Kelley, H. Funakubo, S. Jesse, S. V. Kalinin and M. Ziatdinov, arXiv preprint arXiv:2011.13050 (2020).
37. M. Rashidi and R. A. Wolkow, Acs Nano **12** (6), 5185-5189 (2018).
38. S. V. Kalinin, M. A. Ziatdinov, J. Hinkle, S. Jesse, A. Ghosh, K. P. Kelley, A. R. Lupini, B. G. Sumpter and R. K. Vasudevan, arXiv preprint arXiv:2103.12165 (2021).
39. J. Brehmer and K. Cranmer, arXiv preprint arXiv:2003.13913 (2020).
40. J. Batson, C. G. Haaf, Y. Kahn and D. A. Roberts, arXiv preprint arXiv:2102.08380 (2021).
41. Y. Yu, J. Chen, T. Gao and M. Yu, presented at the International Conference on Machine Learning, 2019 (unpublished).
42. E. Bingham, J. P. Chen, M. Jankowiak, F. Obermeyer, N. Pradhan, T. Karaletsos, R. Singh, P. Szerlip, P. Horsfall and N. D. Goodman, The Journal of Machine Learning Research **20** (1), 973-978 (2019).
43. D. P. Kingma and J. Ba, 2015, arXiv:1412.6980. arXiv.org e-Print archive. https://arxiv.org/abs/1412.6980.
44. H.-S. Daniel, F. J. Breidt and J. v. d. W. Mark, Bayesian Analysis **12** (1), 113-133 (2017).

# SUPPLEMENTARY MATERIALS

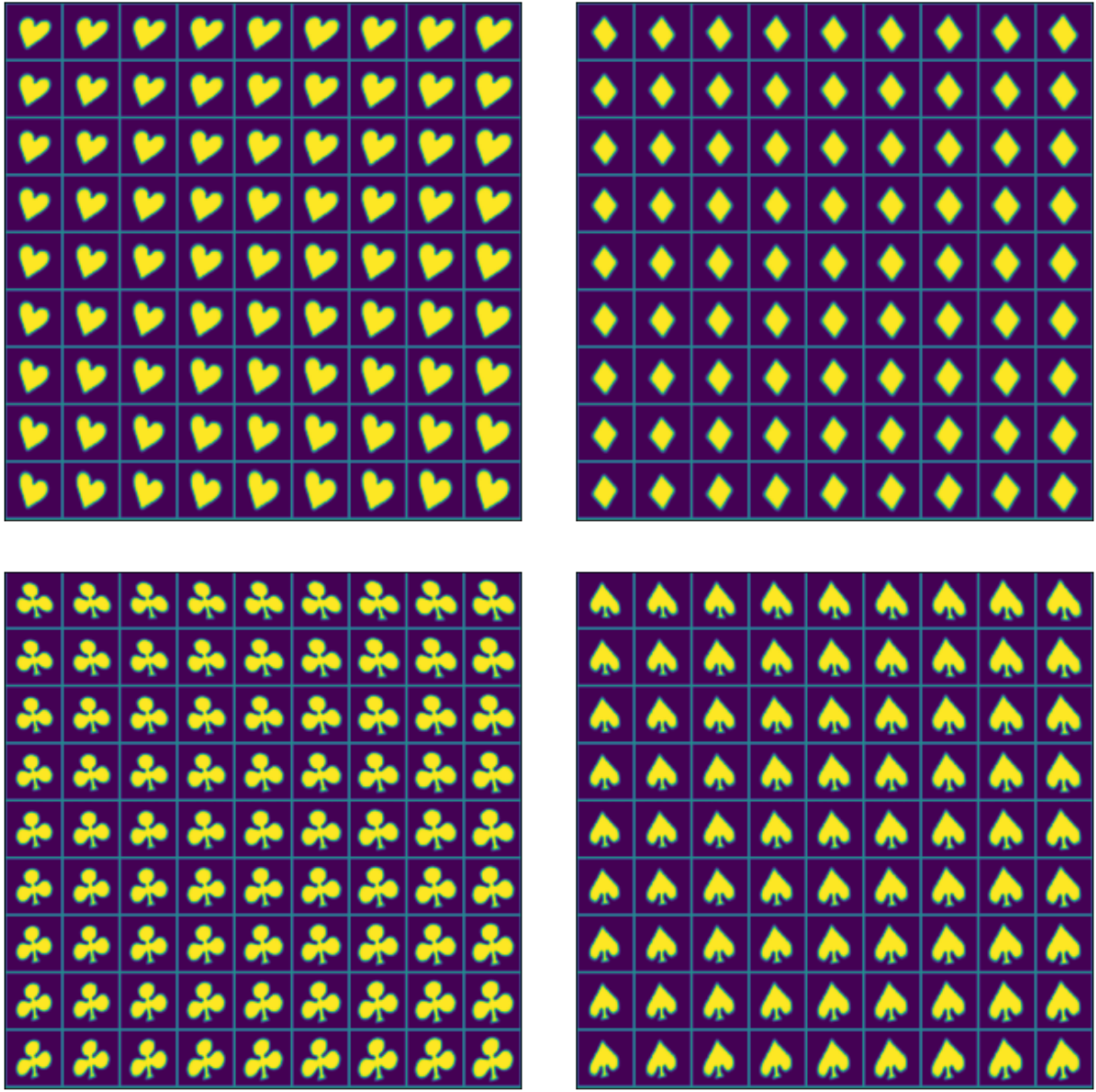

**Figure S1.** Learned latent manifold of *j-tr*VAE for cards dataset with small orientational disorder (see Fig. 1a and Fig. 2b in the main text) shown in the 2D space of continuous latent variables for each inferred discrete class.

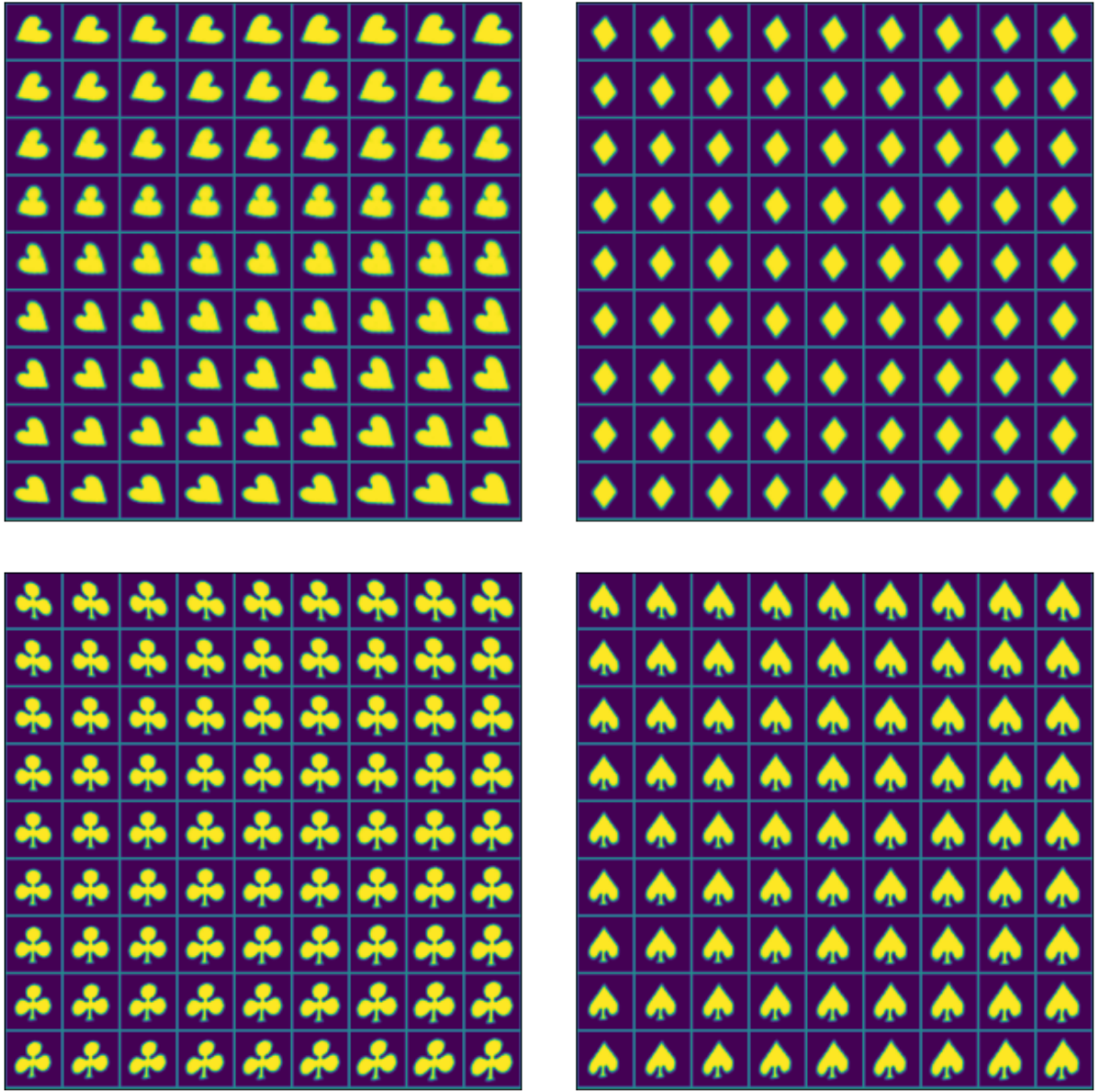

**Figure S2.** Learned latent manifold of *j-tr*VAE for cards dataset with moderate orientational disorder (see Fig. 1b and Fig. 2d in the main text) shown in the 2D space of continuous latent variables for each inferred discrete class.

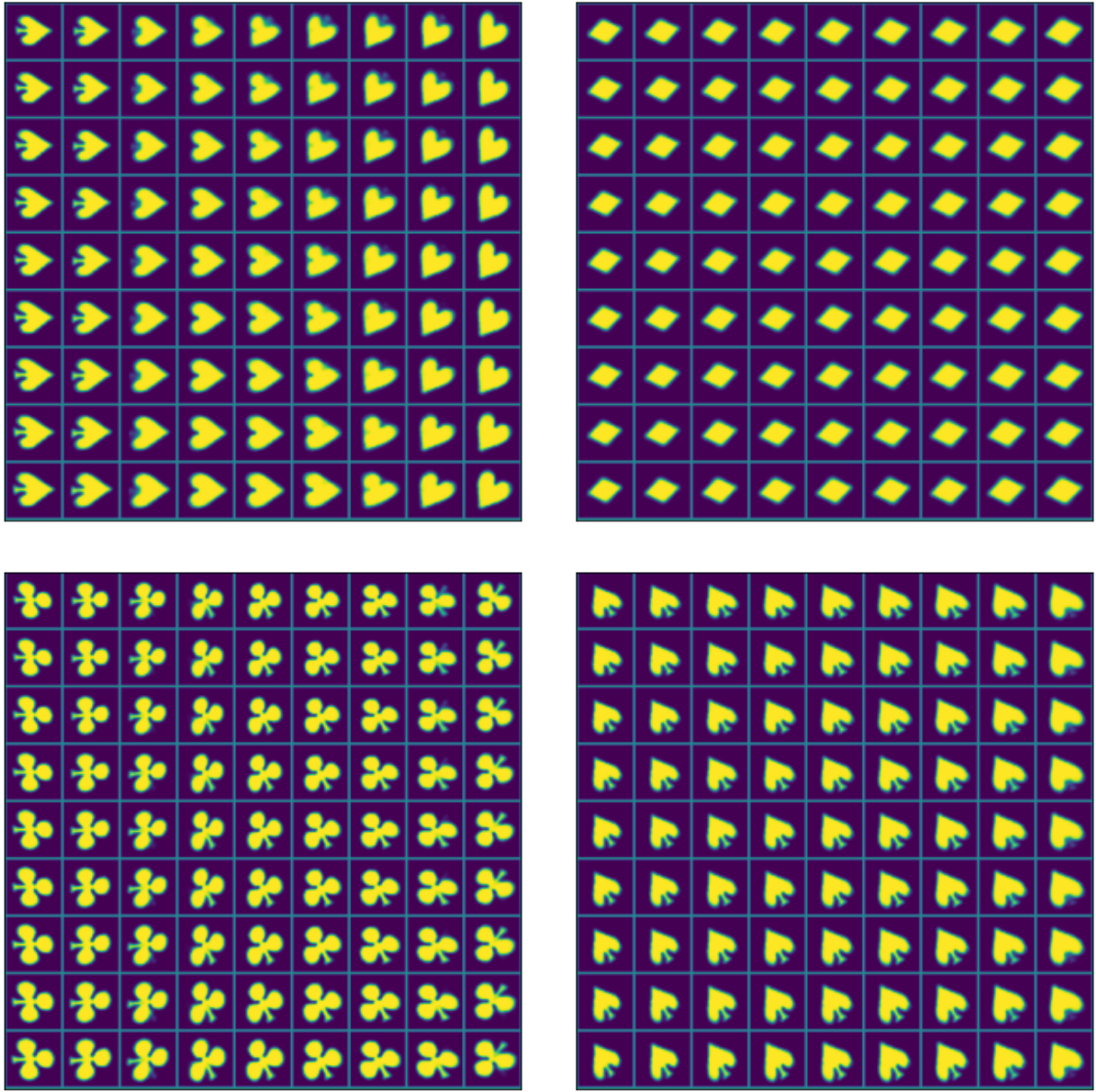

**Figure S3.** Learned latent manifold of *j-tr*VAE for cards dataset with strong orientational disorder (see Fig. 1c and Fig. 2f in the main text) shown in the 2D space of continuous latent variables for each inferred discrete class.

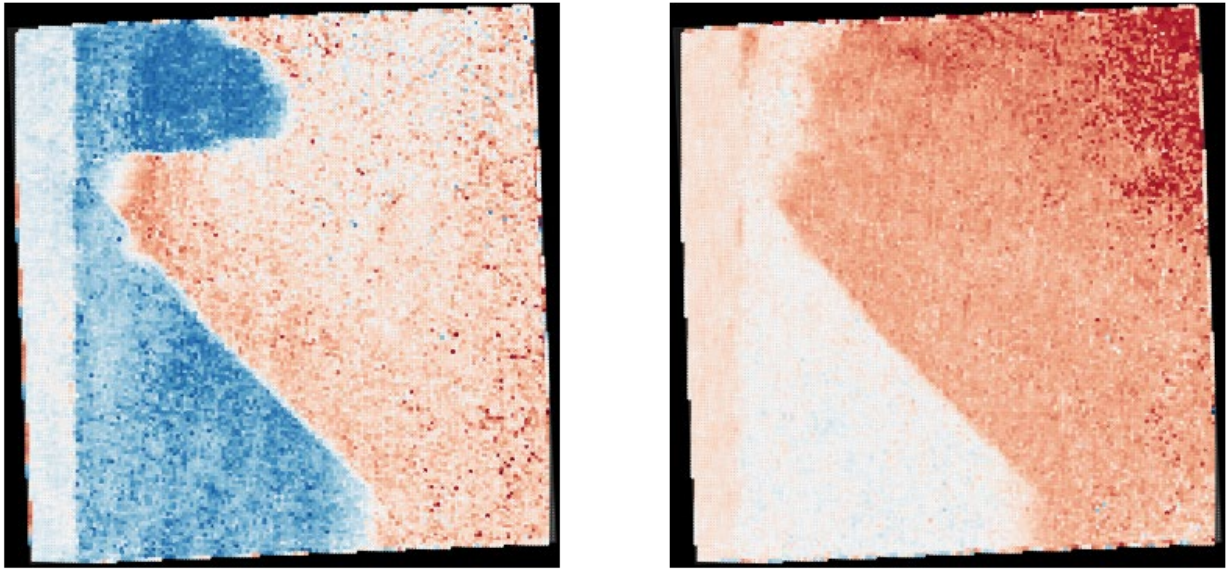

**Fig. S4.** Ground truth polarization values ($y$- and $x$-components) for the BiFeO$_3$ (BFO) on SrTiO$_3$ (001) substrate discussed in the main text.

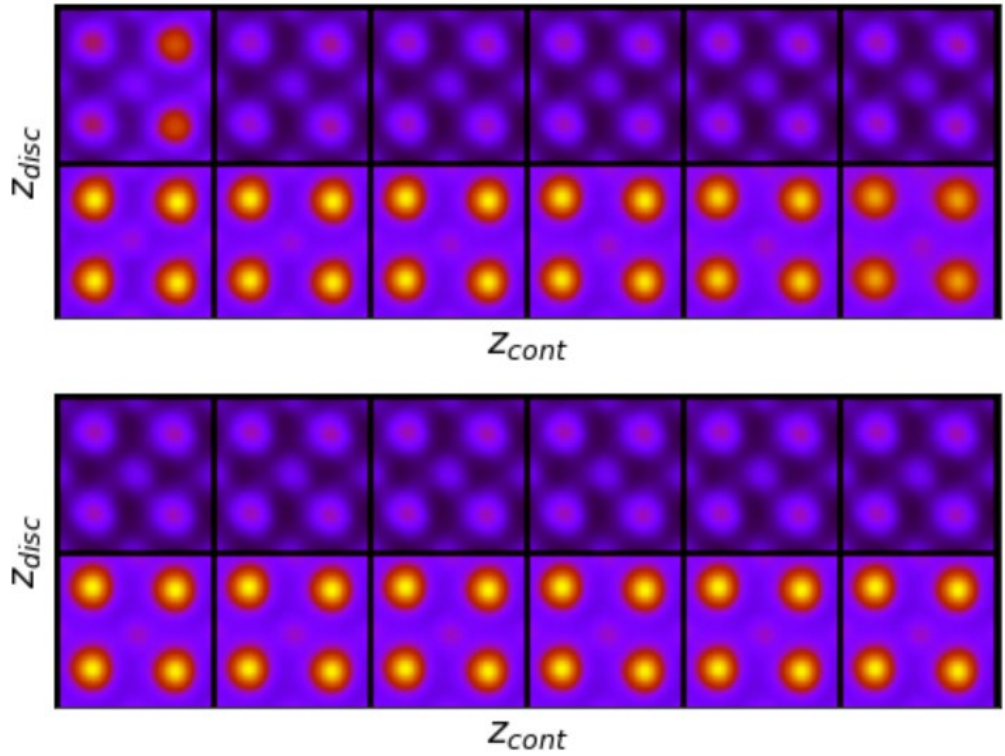

**Fig. S5.** The traversals of the latent manifold for the first and second continuous latent variables learned by *j-tr*VAE from atomically-resolved structural data on BiFeO$_3$ (BFO) on SrTiO$_3$ (001) substrate.

## Supplementary Note 1.

Both fully-connected (MLP) and convolutional architectures can be used with *j-tr*VAE. In both cases, a Cartesian grid of (transformed) coordinates is concatenated with a standard VAE latent vector and fed into the decoder. For the MLP architectures, we use a spatial decoder approach[1] where the decoder model is applied to each coordinate in the grid to reconstruct a full image. For the convolutional architectures, we use a spatial broadcast decoder[2] where the (transformed) coordinates are attached together with a tiled latent vector as additional channels to the convolutional layers in the decoder. We further augmented this architecture by the skip connections from the latent space to each decoder layer to enforce a dependence between the observations and the corresponding latent variables.

**References:**

1. T. Bepler, E. Zhong, K. Kelley, E. Brignole and B. Berger, Advances in Neural Information Processing Systems, 15409-15419 (2019).
2. N. Watters, L. Matthey, C. P. Burgess and A. Lerchner, arXiv preprint arXiv:1901.07017 (2019).